\documentclass[reprint,pra]{revtex4-1}
\usepackage{graphicx}
\usepackage{color}
\usepackage{bm}
\usepackage{amsmath}

\begin{document}

\title{Parametric dependence of bound states in the continuum\\ in
  periodic structures:   vectorial cases}

\author{Lijun Yuan}
\email{ljyuan@ctbu.edu.cn}
\author{Xiaoxia Luo}
\affiliation{College of Mathematics and Statistics, Chongqing Technology and Business University, Chongqing, % 400067,
China \\ Chongqing Key Laboratory of Social Economic and Applied Statistics, Chongqing Technology and Business University, Chongqing, China}

\author{Ya Yan Lu}
%\email{mayylu@cityu.edu.hk}
\affiliation{Department of Mathematics, City University of Hong Kong, 
  Kowloon, Hong Kong, China} 
\date{\today}

\begin{abstract}
A periodic structure sandwiched between two homogeneous media can
support bound states in the continuum (BICs) that are valuable for
many applications.  It is known that generic BICs in periodic structures with an
up-down mirror symmetry and an in-plane inversion symmetry are robust
with respect to structural perturbations that preserve these two
symmetries. For two-dimensional (2D) structures with one periodic
direction and the up-down mirror symmetry (without the in-plane
inversion symmetry), it was recently established  that some scalar BICs can be found by
tuning a single structural parameter. In this paper, we analyze
vectorial BICs in such 2D structures, and show that a typical vectorial BIC
with nonzero wavenumbers in both the invariant and the periodic directions can
only be found by tuning two structural parameters.
Using an all-order perturbation method, we prove that such a
vectorial BIC exists as a curve in the 3D space of three generic
parameters. Our theory is validated by numerical examples involving
periodic arrays of dielectric cylinders. The numerical results also 
illustrate the conservation of topological charge when structural
parameters are varied, if both BICs and circularly polarized states
(CPSs) are included. 
Our study reveals a fundamental property of BICs in periodic structure
and provides a systematically approach for finding BICs in structures
with less symmetry. 
\end{abstract}
\maketitle

\section{Introduction}

Due to their intriguing properties and valuable applications, photonic
bound state in the continuum (BICs) have been extensively studied in
recent years~\cite{hsu16,kosh19,azzam}. In a lossless structure
that is invariant or periodic in one or two spatial directions, a BIC is a
special resonant state with an infinite quality factor ($Q$
factor)~\cite{bonnet94,port05,mari08,plot11,hsu13_2,bulg14,zou15,gomis17},
and it gives high-$Q$ resonances when the structure
or the wavevector is
perturbed~\cite{kosh18,hu18,lijun20,taghi17,rybin17,lijun18pra,jin19,zhen20b,huyuan19}. 
The BICs have found important applications include lasing~\cite{kodi17},
sensing~\cite{yesi19}, harmonic generation~\cite{kosh20},
low-refractive-index light guiding~\cite{zou15,yu19}, etc.
The BICs in periodic structures exhibit interesting topological 
properties. The topological charge of a BIC can be defined using the 
far-field major polarization vector~\cite{zhen14,bulg17pra}. 
If a BIC with a nonzero topological charge 
is destroyed by a perturbation, circularly polarized states (CPSs) 
emerge and the net topological charge remains 
unchanged~\cite{liu20,yoda20,amgad21}.    

Although a BIC typically  becomes a high-$Q$ resonance when the structure
is perturbed~\cite{kosh18,hu18,lijun20}, it may persist if the
perturbation satisfies certain conditions. If a BIC is protected by a
symmetry~\cite{bonnet94,padd00,ochiai01,tikh02,shipman03,lee12}, it
naturally continues to exist when the perturbation preserves the
symmetry. In periodic structures sandwiched between two homogeneous
media, there are propagating BICs with a nonzero
wavevector~\cite{port05,mari08,hsu13_2,bulg14,yang14,gan16}.
These BICs are not protected by symmetry in the usual sense, but they
are robust with 
respect to structural perturbations that maintain the relevant
symmetry~\cite{zhen14,yuan17ol,yuan21}.
Even for perturbations without any symmetry, a BIC is not always
destroyed. In a recent work~\cite{yuan20_2}, we studied 
the continual existence of scalar BICs in periodic
structures without the in-plane inversion symmetry. We showed that for
a generic perturbation with two parameters, a scalar BIC (with a
frequency such that there is 
only one radiation channel) can be maintained by tuning one parameter.
This implies that a typical scalar BIC exists on a curve in the plane
of two generic parameters.

Our previous work is limited to scalar $E$-polarized
BICs~\cite{yuan20_2}. In a 2D structure that is invariant in $x$ and
periodic in $y$, a BIC is associated with wavenumbers $\alpha$ and
$\beta$ corresponding to the $x$ and $y$ directions, respectively.
When $\alpha=0$, the BIC is a scalar one and it can be either
$E$-polarized or $H$-polarized.  The case $\alpha \ne 0$ corresponds
to a vectorial BIC~\cite{bulg17pra}. In this paper, we analyze the
parametric dependence for vectorial BICs in 2D periodic structures. It
turns out that the vectorial BICs for $\beta = 0$ and $\beta \ne 0$
exhibit different behavior in parameter space.  Similar to the scalar
BICs, a vectorial BIC with $\beta=0$ exists as a curve in the plane of two
generic structural parameters. On the other hand, a vectorial BIC with
$\beta \ne 0$ appears as an isolated point in the plane of two
parameters, but exists as a curve in the space of three generic parameters.  This
distinction between the two types of vectorial BICs is somewhat
unexpected, since when the periodic structure has the in-plane
inversion symmetry, they are both robust.

The rest of this paper is organized as follows. In Sec.~\ref{sec:BIC},
we briefly describe the structure, the BICs and related diffraction
solutions. In Sec.~\ref{sec:theory}, we give an outline for a proof
that characterizes the dependence of BICs on structural parameters.
In Sec.~\ref{sec:example}, we present numerical examples to validate
the theory and demonstrate the emergence and annihilation of CPSs when
BICs are destroyed or created. The paper is concluded with some
remarks in Sec.~\ref{sec:conclusion}.

\section{BICs and diffraction solutions}
\label{sec:BIC}

% structure and equations
We consider a 2D lossless dielectric structure that is
invariant in $x$,  periodic in $y$ with period $L$, and symmetric in
$z$ (i.e. the up-down mirror symmetry). 
The dielectric function is real and satisfies  
\begin{equation}
  \label{refper}
  \varepsilon({\bm r}) = \varepsilon(y+L,z)=\varepsilon(y, -z)
\end{equation}
for all ${\bm r} = (y,z)$. It is also assumed that the periodic layer is  sandwiched between two identical homogeneous media with a dielectric constant $\varepsilon_0 \ge 1$. Therefore, 
\begin{equation}
  \label{topbot}
  \varepsilon({\bm r}) = \varepsilon_0, \quad \mbox{for}\  |z| > d, 
\end{equation}
where $d$ is a positive constant.

Since the structure is invariant in $x$, we consider a general eigenmode that depends on $x$ and time $t$ as $\exp[ i (\alpha x - \omega t)]$, where
$\alpha$ is a real wavenumber for the $x$ direction and
$\omega$ is the angular frequency. The electric field is the real part of 
${\bm E}({\bm r}) e^{i (\alpha x - \omega t)}$, where ${\bm E}$  satisfies
 \begin{eqnarray}
 \label{eq:MaxEq1}
&& \left( \nabla + i \alpha {\bm e}_1 \right)  \times  \left(
   \nabla + i \alpha {\bm e}_1 \right)  \times {\bm E} - k^2
   \varepsilon({\bm r}) {\bm E} = 0, \\
&& \label{eq:MaxEq2}
  \left( \nabla + i \alpha {\bm e}_1 \right) \cdot
   [ \varepsilon({\bm r}) {\bm E} ] = 0,
  \end{eqnarray}
  $k = \omega / c$ is the freespace wavenumber, $c$ is the speed
  of light in vacuum,   and ${\bm e}_1 = (1, 0, 0)$ is the unit
  vector in the $x$ direction. Due to the periodicity in $y$, the
  eigenmode can be written as
\begin{equation}
\label{eq:Bloch}
{\bm E({\bm r})} = {\bm \Phi}({\bm r}) e^{i \beta y}, 
\end{equation}
where ${\bm \Phi}({\bm r})$ is periodic in $y$ with period $L$, and $\beta$ is the Bloch wavenumber in $y$ satisfying $|\beta| \le \pi/L$. 

% definition of  BIC  
A BIC is a special guided mode with a real wavevector $(\alpha, \beta)$ and a real frequency $\omega$  satisfying 
$ \sqrt{\alpha^2 + \beta^2 } < k \sqrt{\varepsilon_0}$. 
The electric field of the BIC satisfies ${\bm E}({\bm r}) \to
{\bm 0}$ as $ z \to \pm \infty$ and is normalized such that 
\begin{equation}
\dfrac{1}{L^2} \int_{\Omega} \varepsilon({\bm r}) \overline{ {\bm E}}({\bm r}) \cdot {\bm E}({\bm r}) \, d {\bm r} = 1,
\end{equation}
where $\overline{\bm E}$ is the complex conjugate of ${\bm E}$, and 
$\Omega = \left\{ (y, z) : |y| < L/2,  |z| < +\infty  \right\} $ is one period of the cross section of the structure. 
%
% symmetry of BIC in z
% 
If the BIC is non-degenerate  and its electric field is
${\bm E}({\bm r}) = \left[ E_x({\bm r}),  E_y({\bm r}), E_z({\bm r}) \right]$, we can define a vector field $\tilde{\bm E}$ by 
 \begin{equation}
   \label{eq:Etilde}
   \tilde{{\bm E}}({\bm r}) =
  \left[ E_x(y,-z),  E_y(y,-z), -E_z(y, -z) \right]. 
\end{equation}
Since $\tilde{\bm E}$ is also a BIC for the same frequency and the same
wavevector, we can scale the BIC such that either 
\begin{eqnarray}
&& \label{eq:BICEvenInX} {\bm E}({\bm r}) =
\tilde{{\bm E}}({\bm r}), \quad \mbox{or}  \\
&& \label{eq:BICOddInX} {\bm E}({\bm r}) =
  -\tilde{{\bm E}}({\bm r}).
\end{eqnarray}

In the homogeneous media above or below the periodic layer, plane waves compatible with the BIC have wavevectors $(\alpha, \beta_m, \pm \gamma_m)$, where 
\begin{equation}
\beta_m = \beta + \frac{2\pi m}{L}, \quad 
\gamma_m = \sqrt{ k^2 \varepsilon_0 - \alpha^2 - \beta_m^2}.  
\end{equation} 
In this paper, we only consider BICs satisfying 
\begin{equation}
\label{eq:OneChannel}
\sqrt{\alpha^2 + \beta^2 } < k \sqrt{\varepsilon_0} < \sqrt{ \alpha^2 + \left(  \frac{2 \pi}{L}  - |\beta| \right)^2}.
\end{equation}
The above condition implies that $\gamma=\gamma_0 > 0$ and all other $\gamma_m$ for $m \ne 0$ are pure imaginary. Therefore, the only propagating plane waves are those with wavevectors $(\alpha, \beta, \pm \gamma)$. This ensures that only one radiation channel is open in each side of the periodic layer. 
Moreover, the symmetry condition (\ref{eq:BICEvenInX}) or (\ref{eq:BICOddInX}) implies the radiation channels above and below the periodic layer can be regarded as one channel. 

% diffraction solutions: symmetry in z

In the process of proving our main result, we need diffraction solutions having  the same symmetry in $z$ as the BIC. 
% condition  
 Given the real wavevector ${\bm k} = (\alpha, \beta, \gamma)$, 
we have two real unit vectors ${\bm f}_1$ and ${\bm f}_2$ such that $\{ {\bm k}, {\bm f}_1, {\bm f}_2 \}$ is an orthogonal set. If the BIC satisfies 
condition (\ref{eq:BICEvenInX}),
 we can construct two diffraction solutions 
${\bm E}_1^{(s)}$ and ${\bm E}_2^{(s)}$ that also satisfy condition (\ref{eq:BICEvenInX}). If ${\bm f}_1 = (f_{1x}, f_{1y}, f_{1z})$, then 
${\bm E}_1^{(s)}$ is the solution with incident plane waves
\[
(f_{1x}, f_{1y}, \pm f_{1z}) e^{ i (\beta y \pm \gamma z)}
\]
given below and above the periodic layer (i.e. for $z < -d$ and $z > d$), respectively. Note that we can define a vector field $\tilde{\bm E}^{(s)}_1$ from 
${\bm E}_1^{(s)}$ following Eq.~(\ref{eq:Etilde}), then $( {\bm E}_1^{(s)} +\tilde{\bm E}_1^{(s)} )/2$ solves the same diffraction problem and clearly satisfies condition (\ref{eq:BICEvenInX}). Similarly, using the vector ${\bm f}_2$, we can construct a diffraction solution  ${\bm E}_2^{(s)}$ satisfying condition (\ref{eq:BICEvenInX}).  The case for condition (\ref{eq:BICOddInX}) is similar. 

When there is a BIC, the related diffraction problem does not have a unique solution. For any constant $C_0$, 
${\bm E}_j^{(s)} + C_0{\bm E}$ (for $j=1$, 2) solves the same diffraction problem as ${\bm E}_j^{(s)}$. Without loss of generality, we assume the diffraction solutions are orthogonal with the BIC, i.e., 
 \begin{equation}
 \label{eq:EScatt1_Orth} \int_{\Omega} \varepsilon({\bm r}) \overline{{\bm E}}^{(s)}_j  \cdot {\bm E} \, d {\bm r} = 0, \quad j=1, 2.
 \end{equation}

% for BIC with beta = 0
If the Bloch wavenumber $\beta$ of the BIC is zero, then the vector field given by 
  \begin{equation}
 \label{eq:hatE}
  \hat{{\bm E}}({\bm r}) = \left[ - \overline{E}_x({\bm r}), \overline{E}_y({\bm r}), \overline{E}_z({\bm r})  \right]
  \end{equation}
is also a BIC with the same $\alpha$,  $\beta=0$ and $\omega$. Since we assume the BIC is non-degenerate, there must be a constant $C_1$ such that $\hat{{\bm E}} = C_1 {\bm E}$.
Since the power carried by the BIC is finite, $C_1$ must satisfy $|C_1|=1$. If  $C_1 = e^{ 2 i \varphi}$ for a real $\varphi$, we can replace ${\bm E}$ by $ e^{ i \varphi} {\bm E}$, then the new ${\bm E}$ satisfies
\begin{equation}
\label{eq:RealE} {\bm E}({\bm r})  = \hat{{\bm E}}({\bm r}). 
\end{equation}
This implies that $E_y$ and $E_z$ are real and $E_x$ is pure imaginary. Similarly,  the diffraction solutions ${\bm E}_1^{(s)}$  and ${\bm E}_2^{(s)}$ can also be scaled to satisfy condition (\ref{eq:RealE}). 

\section{Parametric dependence}
\label{sec:theory}

Our objective is to understand how a typical BIC depends on structural
parameters. In general, the dielectric function of a 2D periodic
structure, denoted as $\varepsilon_{\rm g}({\bm r}; {\bm p})$, may
depend on a vector ${\bm p} = (p_1, p_2, ..., 
p_m)$ for $m$ real parameters.
If there is a non-degenerate BIC in the structure when ${\bm p} = {\bm
  p}_*$, we aim to find those ${\bm p}$ near ${\bm p}_*$, such that 
the BIC continues to exist.  
Typically, % a BIC cannot exist for arbitray ${\bm p}$, and  
in the $m$-dimensional parameter space, the values of ${\bm p}$ for
which the BIC exists form a geometric object with a dimension less
than $m$. For scalar $E$-polarized BICs, we have previously proved
that the dimension the geometric object is actually $m-1$, i.e. the
codimension of the geometric object is one~\cite{yuan20_2}.
% This means that BIC parameters
% form a curve in the plane of two parameters (i.e., for $m=2$).
In the following, we show that for vectorial BICs with $\beta=0$ and
$\beta\ne 0$, the codimension is one and two,  respectively.  

For the case of
codimension-$1$, it is sufficient to consider two parameters, i.e.,
$m=2$. In addition, for ${\bm p}$ near ${\bm p}_*$, a general
two-parameter real dielectric function $\varepsilon_{\rm g}({\bm r};
{\bm p})$ can be approximated by  
\begin{equation}
  \label{peps}
\varepsilon({\bm r}) = \varepsilon_* ({\bm r}) 
+ \delta G({\bm r}) + 
\eta F({\bm r}), 
\end{equation}
where $\varepsilon_* ({\bm r}) = \varepsilon_{\rm g}({\bm r}; {\bm p}_*)$, 
${\bm p}_* = (p_{1*}, p_{2*})$, 
$\delta = p_1 - p_{1*}$, $\eta = p_2 - p_{2*}$, 
$G({\bm r})$ and $F({\bm r})$ are partial derivatives of
$\varepsilon_{\rm g}$ (evaluated at ${\bm p}_*$) with respect to $p_1$
and $p_2$, respectively. We assume $\varepsilon_*({\bm r})$ satisfies conditions
(\ref{refper}) and (\ref{topbot}), $F({\bm r})$ and $G({\bm r})$ satisfy condition
(\ref{refper}) and vanish for $|z| > d$.
For the case of codimension-$2$, if $\varepsilon({\bm r})$ is still given by
Eq.~(\ref{peps}), then in general no BICs can be found for $(\delta,
\eta)$ near $(0,0)$. Therefore, we have to introduce three parameters
($m=3$) and replace Eq.~(\ref{peps}) by
\begin{equation}
  \label{peps3}
\varepsilon({\bm r}) = \varepsilon_* ({\bm r}) 
+ \delta G({\bm r}) + \eta F({\bm r}) + s P({\bm r}),
\end{equation}
where $s = p_3 -p_{3*}$, $P({\bm r}) = \partial
\varepsilon_g/\partial p_3 |_{{\bm p} = {\bm p}_*}$, and $P({\bm r})$ satisfies
condition (\ref{refper}) and is zero if $|z| > d$. 
Assuming a BIC exists in the unperturbed structure, we
need to show that for any real $\delta$ 
near zero, there is a real $\eta$ (and a real $s$), such that the BIC
continues its existence in 
the perturbed structure given by Eq.~(\ref{peps}) or
Eq.~(\ref{peps3}). 

We assume the unperturbed structure with the dielectric
function $\varepsilon_*({\bm r})$ has a non-degenerate BIC with
a frequency $\omega_*$, a wavevector $(\alpha_*, \beta_*)$, and  an electric
field ${\bm E}_*({\bm r}) = {\bm \Phi}_*({\bm r}) e^{i \beta_*
  y}$. We also assume the triple $(\alpha_*, \beta_*, \omega_*)$ satisfies condition
(\ref{eq:OneChannel}) and the BIC satisfies Eq.~(\ref{eq:BICEvenInX}).
Let the diffraction solutions corresponding to the BIC, as introduced
in Sec.~II, be ${\bm E}_j^{(s)}({\bm r}) = {\bm \Psi}_j({\bm r})
e^{ i \beta_* y}$ for $j=1$, 2. We assume they satisfy
Eqs.~(\ref{eq:BICEvenInX}) and (\ref{eq:EScatt1_Orth}).
For the perturbed structure given by Eq.~(\ref{peps}) or
Eq.~(\ref{peps3}), we look for a BIC, near the one in the
unperturbed structure, with a frequency $\omega$,   a wavevector $(\alpha, \beta)$, 
and electric field ${\bm E}({\bm r})  = {\bm \Phi}({\bm r}) e^{ i \beta y}$.

First, we establish a codimension-$2$ result for vectorial BICs with
$\alpha_* \ne 0$ and $\beta_* \ne 0$.  For a perturbed structure with
$\varepsilon({\bm r})$ given in Eq.~(\ref{peps3}), a BIC, if it
exists, can be found by expanding ${\bm \Phi}$, $\alpha$, $\beta$,  $k$,
$\eta$ and $s$ in power series of $\delta$:  
\begin{eqnarray}
  \label{series1}
{\bm \Phi} &=& {\bm \Phi}_* + {\bm \Phi}_1 \delta + {\bm \Phi}_2 \delta^2 + \cdots \\
\label{series2a}
\alpha &=& \alpha_* + \alpha_1 \delta + \alpha_2 \delta^2 + \cdots\\ 
  \label{series2}
  \beta &=& \beta_* + \beta_1 \delta + \beta_2 \delta^2 + \cdots\\ 
  \label{series3}
k &=& k_* + k_1 \delta + k_2 \delta^2 + \cdots \\
  \label{series4}
  \eta &=&  \quad \quad   \eta_1 \delta + \eta_2 \delta^2 + \cdots \\
  \label{series5}
s &=&  \quad \quad   s_1 \delta + s_2 \delta^2 + \cdots   
\end{eqnarray}
Here, $\delta$ is a small and arbitrary real number, $\eta$ and $s$ 
depend on $\delta$ and are also expanded. Inserting  the above into
the governing equations for ${\bm \Phi}$, i.e., Eqs.~(\ref{eq:MaxEq3}) and
(\ref{eq:MaxEq4}) in the Appendix,  and
comparing the coefficients of $\delta^j$ for each $j \ge 1$, we obtain
the following differential equation for ${\bm \Phi}_j$:
\begin{eqnarray}
  \label{eq:Phij}
 && \mathcal{L} {\bm \Phi}_j = \alpha_j \mathcal{B}_1 {\bm \Phi}_* + \beta_j
                        \mathcal{B}_2 {\bm \Phi}_*
                        + 2 k_* k_j \varepsilon_* {\bm \Phi}_* \nonumber \\
  && \qquad
     + k_*^2 \eta_j F({\bm r}) {\bm \Phi}_* 
          + k_*^2 s_j P({\bm r}) {\bm \Phi}_* + {\bm C}_j({\bm r}), \\
&& \label{eq:Phij2} (\nabla + i {\bm \alpha}_*) \cdot (\varepsilon_*
   {\bm \Phi}_j)
   =  \alpha_j h_1({\bm r}) + \beta_j h_2({\bm r}) \nonumber \\
  &&\qquad + \eta_j h_3({\bm r}) + s_j h_4({\bm r}) + g_j({\bm r}),
\end{eqnarray}
where
${\bm \alpha}_* = (\alpha_*, \beta_*, 0)$, 
${\cal L}$, ${\cal B}_1$, ${\cal B}_2$ are differential
operators,  
$h_m$ (for $m=1$, 2, 3, 4) are scalar functions depending
on $\alpha_*$, $\beta_*$, $k_*$ and $\varepsilon_*({\bm r})$,
${\bm C}_j$  is a vector field  and $g_j$ is a
  scalar function depending on all 
previous iterations such as ${\bm \Phi}_m$ for $m < j$. The details are
given in the Appendix. Importantly,  the BIC exists if
and only if,  {\em for each $j \geq 1$, $\alpha_j$, $\beta_j$,
$k_j$, $\eta_j$ and $s_j$ can be solved and they are real, and ${\bm \Phi}_j$ can be solved from
Eqs.~(\ref{eq:Phij}) and (\ref{eq:Phij2}) and it decays exponentially  to zero as $|z| \to
\infty$.}

Integrating the dot products of Eq.~(\ref{eq:Phij}) with the complex
conjugates of ${\bm \Phi}_*$, ${\bm \Psi}_1$ and ${\bm \Psi}_2$ on $\Omega$, 
we obtain a $3\times 5$ linear system 
\begin{equation}
  \label{3by5} 
  \left[ \begin{matrix} a_{11} & a_{12} & a_{13} & a_{14}  & a_{15} \cr 
a_{21} & a_{22} & 0 & a_{24} & a_{25} \cr 
a_{31} & a_{32} & 0 & a_{34}  & a_{35} \end{matrix} \right]
\left[ \begin{matrix} \alpha_j \cr \beta_j \cr  k_j \cr \eta_j \cr s_j 
    \end{matrix} \right] = 
\left[ \begin{matrix} {b}_{1j} \cr {b}_{2j} \cr 
    {b}_{3j} \end{matrix} \right]
\end{equation}
with a $3\times 5$ coefficient matrix ${\bm   A}$  and  a right hand
side ${\bm b}_j$. The entries  of ${\bm A}$ and ${\bm b}_j$ are given
in the Appendix. While 
Eq.~(\ref{3by5}) is a necessary condition for Eq.~(\ref{eq:Phij}) to
have a solution that decays exponentially to zero as $|z| \to \infty$,
it is also a sufficient condition. More precisely, if all previous
iterations ${\bm \Phi}_m$ for $m < j$ decay to zero exponentially as $|z|
\to \infty$,  and $(\alpha_j, \beta_j, k_j, \eta_j, s_j)$ is a real solution of Eq.~(\ref{3by5}), then
Eq.~(\ref{eq:Phij}) always has a solution that decays to zero
exponentially as $|z| \to \infty$.
It is easy to show that all 
entries in the first row of matrix ${\bm A}$ and 
$b_{1j}$ are real, thus, Eq.~(\ref{3by5}) is equivalent to the 
following real $5\times 5$ linear system: 
\begin{equation}
\label{5by5}  
 \left[ \begin{matrix} a_{11} & a_{12} & a_{13} & a_{14}  & a_{15} \cr 
a'_{21}  & a'_{22} & 0 & a'_{24} & a'_{25}  \cr 
a'_{31}  & a'_{32} & 0 & a'_{34}  & a'_{35} \cr 
a''_{21}  & a''_{22} & 0 & a''_{24} & a''_{25}  \cr 
a''_{31}  & a''_{32} & 0 & a''_{34}  & a''_{35} 
 \end{matrix} \right] 
\left[ \begin{matrix} \alpha_j \cr \beta_j  \cr k_j\cr \eta_j  \cr s_j \end{matrix} \right]
= \left[ \begin{matrix} {b}_{1j} \cr b'_{2j} \cr b'_{3j} \cr b''_{2j} \cr b''_{3j} \end{matrix} \right], 
\end{equation}
where $\xi'$ and $\xi''$ denote the real and imaginary parts of any
complex number $\xi$. The real $5 \times 5$
coefficient matrix above is related to the BIC (of the unperturbed structure), the corresponding diffraction solutions, and the perturbation 
profiles $F({\bm r})$ and $P({\bm r})$. If this $5\times 5$ matrix is
invertible, then for each $j\ge 1$, $(\alpha_j, \beta_j, k_j, \eta_j, s_j)$ can be solved from Eq.~(\ref{5by5}) and it is real, ${\bm \Phi}_j$ can be solved from
Eq.~(\ref{eq:Phij}) and it decays to zero exponentially as $|z| \to
\infty$. This implies that a BIC exists in the perturbed structure
with $\eta$ and $s$ depending on $\delta$ and given by
Eqs.~(\ref{series4}) and (\ref{series5}), respectively.

To establish  a possible codimension-$1$ result, we assume $\varepsilon({\bm 
  r})$ is given by Eq.~(\ref{peps}) and expand ${\bm \Phi}$, $\beta$,
$\alpha$, $k$ and $\eta$ as in
Eqs.~(\ref{series1})-(\ref{series4}). This leads to a slightly
simplified version of Eq~(\ref{eq:Phij}) without terms involving 
$P({\bm r})$ and $s_j$ in the right hand and inside ${\bm C}_j$.  
A necessary and sufficient condition for this simplified Eq.~(\ref{eq:Phij}) to
have an exponentially decaying solution is
\begin{equation}
  \label{3by4} 
  \left[ \begin{matrix} a_{11} & a_{12} & a_{13} & a_{14}  \cr 
a_{21} & a_{22} & 0 & a_{24}  \cr 
a_{31} & a_{32} & 0 & a_{34}  \end{matrix} \right]
\left[ \begin{matrix} \alpha_j \cr   \beta_j \cr k_j \cr \eta_j  
    \end{matrix} \right] = 
\left[ \begin{matrix} {b}_{1j} \cr {b}_{2j} \cr 
    {b}_{3j} \end{matrix} \right].
\end{equation}
The first row of the above $3\times 4$ coefficient matrix and $b_{11}$ are real, but for a general vectorial
BIC with $\alpha_* \ne 0$ and $\beta_* \ne 0$, the second and third
rows of the coefficient matrix are complex. Therefore, Eq.~(\ref{3by4}) is
equivalent to a real system with five equations and 
four unknowns, and in general, it does not have a real solution. This
means that (in general) there is no BIC in the perturbed structure with a small
$\delta$ and small $\eta$. 

For a vectorial BIC with $\beta_* = 0$ and $\alpha_* \ne 0$,
we have shown (in Sec.~II) that the electric field ${\bm E}_*$ of
the BIC and the
corresponding diffraction solutions ${\bm E}_1^{(s)}$ and ${\bm E}_2^{(s)}$
can be scaled to satisfy Eq.~(\ref{eq:RealE}). This implies that 
$a_{22}$ and $a_{32}$ are pure imaginary, all other elements of the
$3\times 4$ matrix in Eq.~(\ref{3by4}) are real. Therefore, if $a_{21}
a_{34} - a_{24} a_{31} \ne 0$,  then for each $j 
\ge 1$, the linear system (\ref{3by4}) has a solution with $\beta_j = 0$ and real ($\alpha_j$, $k_j$, $\eta_j$) satisfying
\begin{equation}
  \label{3by3}
 \left[ \begin{matrix} a_{11} & a_{13} & a_{14} \cr 
 a_{21} & 0 & a_{24} \cr 
 a_{31} & 0 & a_{34} \end{matrix} \right] 
\left[ \begin{matrix}  \alpha_j \cr k_j\cr \eta_j \end{matrix} \right]
= \left[ \begin{matrix} b_{1j} \cr b_{2j} \cr 
b_{3j} \end{matrix} \right]. 
\end{equation}
The result is established recursively with 
additional details given in the Appendix. 
The key step is to show that $b_{1j}$, $b_{2j}$ and $b_{3j}$ are
real, then the complex linear system (\ref{3by4}) gives $\beta_j = 0$ 
and the real linear system (\ref{3by3}). The matrix entry $a_{13}$ is
always nonzero. The condition $a_{21} a_{34} - a_{24} a_{31} \ne 0$ ensures that the
coefficient matrix in linear system (\ref{3by3}) is invertible. The
matrix entries $a_{21}$, $a_{24}$, $a_{31}$ and $a_{34}$ are related to ${\bm \Phi}_*$ (the BIC
${\bm E}_*$), ${\bm \Psi}_1$ (the diffraction solution ${\bm E}_1^{(s)}$),
${\bm \Psi}_2$ (the diffraction solution ${\bm E}_2^{(s)}$), and the
perturbation profile $F({\bm r})$. Therefore, the BIC and the
perturbation profile $F({\bm r})$ must satisfy the extra condition
$a_{21} a_{34} - a_{24} a_{31} \ne 0$. Notice that no extra condition on
perturbation profile $G$ is needed. 

In Ref.~\cite{yuan20_2},  we analyzed the the parametric dependence of
scalar $E$-polarized BICs. Our current formulation is applicable to 
both $E$- and $H$-polarized scalar BICs. If $\alpha_*=0$ and $\beta_*
\ne 0$, the BIC is a scalar one with either the $E$ or $H$ polarization. 
  For example, if the BIC is $H$-polarized, we can choose
  ${\bm f}_1$ and ${\bm f}_2$ such that ${\bm E}_1^{(s)}$ and
  $ {\bm E}_2^{(s)}$ are $E$- and $H$-polarized, respectively.
  It can be easily shown that 
$a_{11}=a_{22} = a_{31}=a_{24}=0$ and $b_{2j} = 0$ for all $j \geq 1$.
As we mentioned earlier, 
all entries in the first row of ${\bm A}$ and $b_{1j}$ are real. If
$a_{32} \neq 0 $ and $\mbox{Im}(a_{34}/a_{32}) \neq 0 $,
then the linear system
(\ref{3by4}) has a real solution with $\alpha_j = 0$
and real $\beta_j$, $k_j$ and $\eta_j$ satisfying 
\begin{equation}
\label{2by3}  
 \left[ \begin{matrix} a_{12} & a_{13} & a_{14} \cr 
 a'_{32} & 0 & a'_{34} \cr 
 a''_{32} & 0 & a''_{34} \end{matrix} \right] 
\left[ \begin{matrix}  \beta_j \cr k_j\cr \eta_j \end{matrix} \right]
= \left[ \begin{matrix} b_{1j} \cr b'_{3j} \cr 
b''_{3j} \end{matrix} \right]. 
\end{equation}
More details are given in the Appendix. 

If the BIC is a standing wave, i.e., $\alpha_* = \beta_* = 0$, we can
show that $a_{11} = a_{22} = a_{31}=a_{24}=0$, $a_{32}$ is pure
imaginary, all other elements of ${\bm A}$ are real, and 
if $a_{34} \neq 0$, then the linear system (\ref{3by4})
has a real solution with $\beta_j=\alpha_j=0$, and $k_j$ and $\eta_j$
satisfying 
\begin{equation}
\label{2by2}  
 \left[ \begin{matrix}  a_{13} & a_{14} \cr 
  0 & a_{34} \end{matrix} \right] 
\left[ \begin{matrix} k_j\cr \eta_j \end{matrix} \right]
= \left[ \begin{matrix} b_{1j} \cr 
b_{3j} \end{matrix} \right]. 
\end{equation}
It can be shown  recursively that for all $j\ge 1$, $b_{2j}=0$, and $b_{1j}$ 
and $ b_{3j}$ are real.

In previous works~\cite{yuan17ol,yuan21},  we analyzed the
robustness of BICs in periodic structures with both up-down mirror
symmetry and in-plane inversion symmetry. It has been shown that a
generic BIC in a periodic structure with these 
two symmetries continues its existence when the structure is perturbed
preserving both symmetries. For 2D structures, the
in-plane inversion symmetry is simply the reflection symmetry in $y$,
i.e,  $\varepsilon({\bm r}) = \varepsilon(-y,z)$ 
for all ${\bm r}$. Actually, the existing robustness result for 2D structures (with 1D
periodicity) covers only the scalar $E$-polarized
BICs~\cite{yuan17ol}. In the following, we briefly establish the robustness
for both scalar and vectorial BICs. Let $\varepsilon_*({\bm r})$ be
the dielectric function of the unperturbed structure as before. We
consider a perturbed structure with a dielectric function given by 
\begin{equation}
  \label{peps1}
  \varepsilon({\bm r}) =   \varepsilon_*({\bm r}) + \delta G({\bm r}),
\end{equation}
where $\varepsilon_*({\bm r})$ and $G({\bm r})$ satisfy the conditions
stated earlier in this section, and are even functions of
$y$. As in previous works~\cite{yuan17ol,yuan21}, we can show
that the BIC in the unperturbed structure can be scaled such that the
$x$ and $y$ components of its electric field ${\bm E}_*$ are ${\cal
  PT}$-symmetric and the $z$ component is anti-${\cal PT}$-symmetric,
namely, 
    \begin{equation}
 \label{eq:W}
  {\bm E}_*({\bm r}) = 
  \left[  \overline{E}_{*x}(-y,z), \overline{E}_{*y}(-y,z), -
    \overline{E}_{*z}(-y,z) \right],
\end{equation}
where $\overline{E}_{*x}$ is the complex conjugate of $E_{*x}$, etc. 
Meanwhile, the two diffraction solutions ${\bm E}_1^{(s)}$ and ${\bm E}_2^{(s)}$ can also be scaled to have ${\cal PT}$-symmetric $x$ and
$y$ components and anti-${\cal PT}$-symmetric $z$ components. 
Following the same expansions (\ref{series1})-(\ref{series3}), we
obtain a simplified version of Eq.~(\ref{eq:Phij}) and the following
linear system:
  \begin{equation}
\label{Symm3by3}  
 \left[ \begin{matrix} a_{11} & a_{12} & a_{13} \cr 
 a_{21} & a_{22} & 0 \cr 
 a_{31} & a_{32} & 0 \end{matrix} \right] 
\left[ \begin{matrix}  \alpha_j \cr \beta_j \cr k_j \end{matrix} \right]
= \left[ \begin{matrix} b_{1j} \cr b_{2j} \cr 
b_{3j} \end{matrix} \right], 
\end{equation}
where the coefficient matrix consists of the first three columns of
matrix ${\bm A}$. The additional symmetries in $y$ (for the structure,
the perturbation, the BIC and the diffraction solutions) allow us to show
that all entries of this $3 \times 3$ matrix are real. Since $a_{13}$ is
always nonzero, if $a_{21} a_{32} - a_{22} a_{31} \neq 0$, the above 
$3\times 3$ matrix is invertible. Under this condition, we can show that for
each $j \ge 1$, the vector ${\bm b}_j$ in the right hand side is real,
$(\alpha_j, \beta_j, k_j)$ can be solved and is real, ${\bm \Phi}_j$ can be
solved from (the simplified) Eq.~(\ref{eq:Phij}) and it decays to zero
exponentially as $|z| \to \infty$. Therefore, a BIC exists in the
perturbed structure given by Eq.~(\ref{peps1}).

In summary, we have studied how BICs in 2D structures (with 1D
periodicity) exist continuously under small structural
perturbations. For structures without the reflection symmetry in $y$,
a vectorial BIC with $\alpha_* \ne 0 $ and $\beta_* \ne 0$ can exist
continuously by adding two additional parameters (i.e, $\eta$ and $s$
depending on $\delta$), all other BICs (with $\alpha_*=0$ and/or
$\beta_*=0$) can exist continuously by adding one additional parameter
(i.e. $\eta$ as a function of $\delta$). For structures with the
reflection symmetry in $y$, these BICs exist continuously with the
perturbation (i.e., they are robust). These conclusions are obtained
only for non-degenerate BICs satisfying the single-radiation-channel condition
(\ref{eq:OneChannel}). 
The up-down mirror symmetry is always
assumed so the the radiation channels below and above the layer
(i.e. for $z < -d$ and $z > d$) are exactly the same. The BIC and the
perturbation profiles must satisfy additional generic conditions so
that the coefficient matrix of a linear system, such as
Eq.~(\ref{5by5}), is invertible.

\section{Numerical examples}
\label{sec:example}
To validate our theory, we consider BICs in periodic 
arrays of dielectric cylinders (with dielectric constant
$\varepsilon_1$) surrounded by air (with dielectric constant
$\varepsilon_0 = 1$). The original unperturbed structure consists of
circular cylinders with radius $a$ and is symmetric in both $y$ and
$z$ directions.
BICs in a periodic arrays of circular cylinders have been studied by
many authors~\cite{shipman03,bulg14,bulg17pra}. In Table~\ref{table:BIC},
\begin{table}[htp]
\centering 
\begin{tabular}{c|c|c|c|c|c|c} \hline 
BICs & \ $a/L$ \ & \ $\varepsilon_1$ \  & ${\alpha_* L}/({2\pi})$ & ${\beta_* L}/(2\pi)$ & ${\omega_* L}/(2\pi c)$  & $q$ \\ \hline 
S1 & $0.3 $ & $10$ & $0$ & $ 0$ & $0.4414$ & $-1$  \\ \hline 
S2 & $0.3 $ & $10$ & $0$ & $ 0.2206$ & $0.6174$ & $+1$  \\ \hline 
V1 & $0.44 $ & $15$ & $0.3068$ & $ 0$ & $0.5862$ & $-1$  \\ \hline 
V2 & $0.3 $ & $10$ & $ 0.2931$ & $0.1774$  & $0.6180$ & $-1$ \\ \hline 
\end{tabular}
\caption{Scalar and vectorial BICs in periodic arrays of circular
  cylinders (surrounded by air) with radius $a$ and dielectric constant $\varepsilon_1$.}
\label{table:BIC}
\end{table}
we list four BICs for different structure parameters. The topological charges
of the BICs, defined using the far field major polarization vector of the
resonant modes~\cite{zhen14,bulg17pra,yoda20}, are also listed in the
table. 

%The wave fields (i.e. $z$-component of ${\bm \Phi}_{*}$) of above BICs are shown in Fig.~(\ref{fig:Field}).
%\begin{figure}[htp]
%\centering
%\includegraphics[scale=0.5]{./figs/FigPolarizationDirection3.eps}
%\includegraphics[scale=0.5]{./figs/FigPolarizationDirection3.eps} \\
%\includegraphics[scale=0.5]{./figs/FigPolarizationDirection3.eps}
%\includegraphics[scale=0.5]{./figs/FigPolarizationDirection3.eps}
%\caption{Fields of BICs. (a) - (d): wave field (i.e. $z$-component of ${\bm \Phi}_{*}$) of BIC1 to BIC4, respectively.}
%\label{fig:Field}
%\end{figure}

\begin{figure}[htp]
\centering
\includegraphics[scale=0.9]{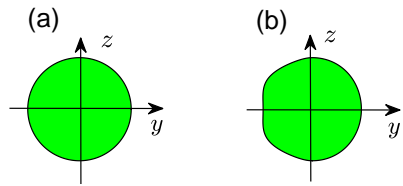}
\vspace*{-0.6cm}
\caption{Cross sections of the original and perturbed cylinders. (a): the original circular cylinders for $a=0.3L$. (b): the perturbed non-circular cylinders for $a = 0.3 L$ and  $(\delta, \eta) = (0.05, 0.05)$.}
\label{fig:Structure}
\end{figure}

The perturbed structure consists of cylinders with the boundary (of the
cylinder centered at the origin) given by 
\begin{equation}
\label{eq:Perturbation}
 y = \left\{ \begin{matrix}  a \left[ \sin(\theta) - \delta \sin^2(\theta) \right], & \theta \in [0, \pi), \\
a \left[ \sin(\theta) - \eta \sin(5\theta) \right], & \theta \in [\pi, 2\pi),
 \end{matrix}  \right.  \ \ z = a \cos(\theta) 
\end{equation}
where $\delta$ and $\eta$ are real parameters. 
Note that for $(\delta, \eta) \neq (0,0)$, the structure is symmetric
in $z$ and not symmetric in $y$. In Fig.~\ref{fig:Structure}, we show
the cross sections of original and perturbed cylinders, respectively,  for $a=0.3L$,
$\delta = 0.05$ and $\eta = 0.05$. 

\begin{figure}[htp]
\centering 
\includegraphics[scale=1]{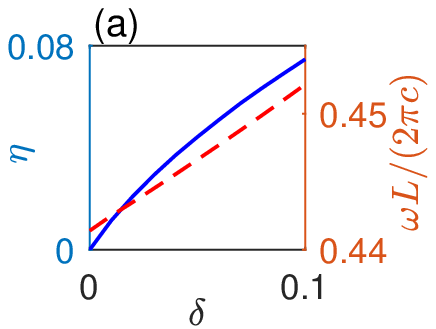}
\includegraphics[scale=1]{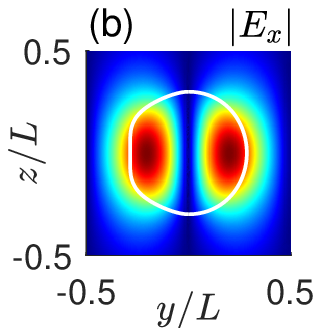}
\includegraphics[scale=1]{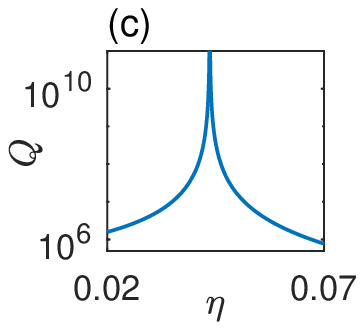}
\caption{BIC near S1 in a perturbed array of cylinders. (a): parameter 
  $\eta$ (solid blue curve, left axis) and frequency $\omega$ (dashed red curve, right axis) of the BIC 
  for different values of $\delta$. (b): 
  magnitude of the electric field  (i.e. $|E_x|$) of the BIC for $\delta=0.05$ and $\eta = 0.0440$. (c): $Q$
  factor of resonant modes (with $\alpha=\beta=0$) as a function of 
  $\eta$ for $\delta = 0.05$.  The peak at $\eta = 0.0440$ corresponds to a BIC with $\omega = 0.4467 (2\pi c/L)$. } 
\label{fig:BIC1}
\end{figure}

The BIC S1 is a scalar $E$-polarized standing wave ($\alpha_*=\beta_*=0$). Our
theory predicts that it exists continuously on a curve in the
$\delta$-$\eta$ plane. 
This is confirmed by numerical results for $\delta \in [0,
0.1]$. The BIC remains as a standing wave, i.e., $\alpha=\beta=0$,  for all $\delta$. 
The parameter $\eta$ and
the frequency $\omega$ of the BIC depend on $\delta$, and are shown 
in Fig.~\ref{fig:BIC1}(a)
as the solid blue and dashed red curves,
respectively. For $\delta = 0.05$ and $\eta = 0.0440$, the perturbed structure
has a BIC with $\omega  =  0.4467 (2\pi c/L)$. The magnitude
of the electric field (i.e. $|E_x|$) of this BIC is shown in Fig.~\ref{fig:BIC1}(b).
In Fig.~\ref{fig:BIC1}(c),  we show the 
$Q$ factor of resonant modes (with $\alpha=\beta=0$) for fixed 
$\delta = 0.05$ and different $\eta$.

\begin{figure}[htp]
\centering 
\includegraphics[scale=0.8]{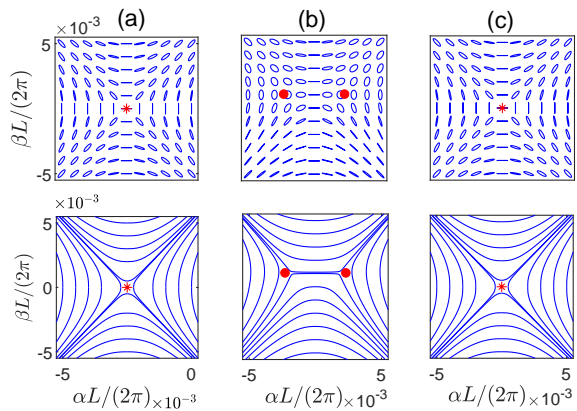}
\caption{Emergence and annihilation of CPSs near BIC S1. (a), (b) and (c): polarization ellipses (top panels) and polarization directions (lower panels) for  $(\delta,\eta) = (0,0)$, $(0.05, 0)$ and $(0.05, 0.0440)$, respectively. Red asterisks and dots  denote BICs and CPSs, respectively.}
\label{fig:PolarizationBIC1}
\end{figure}

Circularly polarized resonant states (CPSs) and BICs are polarization
singularities in the momentum space (the $\alpha$-$\beta$ plane)
\cite{zhen14,bulg17pra,liu20,yoda20}. In
Fig.~\ref{fig:PolarizationBIC1},
we show polarization ellipses
(upper panels) and
polarization directions (lower panels) for resonant modes in 
periodic arrays of cylinders with (a) $\delta=\eta=0$, (b)
$\delta=0.05$ and $\eta=0$, and (c) $\delta=0.05$ and $\eta =
0.0440$. The resonant modes are from the band that contains BIC S1 when
$\delta=\eta=0$. The BIC S1 and its extension in a perturbed
array are shown as the red  asterisks in panels (a) and (c),
respectively. For fixed $\eta=0$ and when $\delta$ is increased, BIC
S1 is destroyed and turned to a pair of  CPSs 
with topological charge $-1/2$. The CPSs are shown as red dots in
Fig.~\ref{fig:PolarizationBIC1}(b). It can be easily shown that the 
two CPSs must have the same $\beta$ and opposite $\alpha$.
% Due to this property, as long as one of the two CPSs is
% moved on the $\beta$-axis, the two CPSs must merge on the $\beta$-axis
% and generate a BIC.
For a fixed $\delta=0.05$ and when $\eta$ is increased, the two CPSs
move towards the $\beta$ axis, and form a BIC at $\eta = 0.0440$
as shown in Fig.~\ref{fig:PolarizationBIC1}(c).

\begin{figure}[htp]
\centering 
\includegraphics[scale=1]{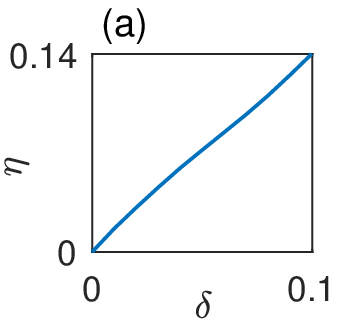}
\includegraphics[scale=1]{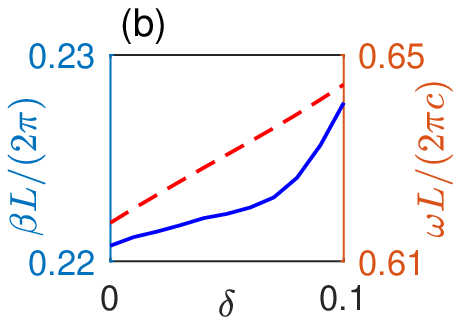}
\includegraphics[scale=1]{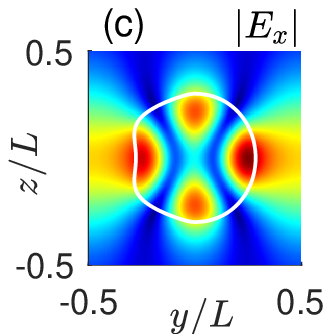}
\includegraphics[scale=1]{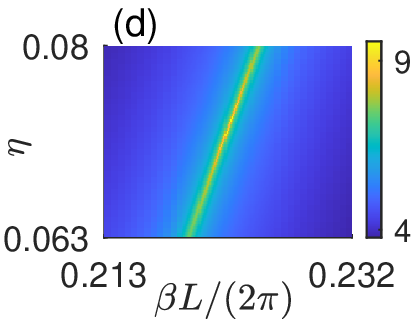}
\caption{BIC near S2 in a perturbed array of cylinders. (a): parameter $\eta$ of the BIC as a function of $\delta$.  (b): wavenumber $\beta$ (solid blue curve, left axis)  and frequency $\omega$ (dashed red curve, right axis) of the BIC as functions of $\delta$.  (c): magnitude of the BIC (i.e. $|{E}_x|$) for $\delta = 0.05, \eta = 0.0721$, $\alpha = 0, \beta = 0.2223 (2\pi/L)$, and $\omega = 0.6307 (2\pi c/L)$.  (d): logarithmic value of the $Q$ factor (i.e. $\log_{10}Q$) of resonant modes as a function of $\beta$ and $\eta$ for $\alpha=0$ and $\delta = 0.05$. }
\label{fig:BIC2}
\end{figure}

\begin{figure}[htp]
\centering 
\includegraphics[scale=0.8]{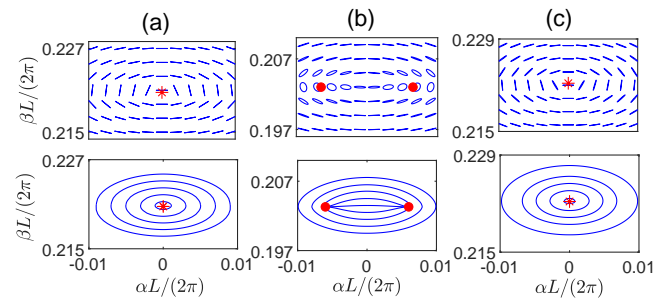}
\caption{Emergence and annihilation of CPSs near BIC S2. (a), (b) and
  (c): polarization ellipses (top panels) and polarization directions
  (lower panels) for  $(\delta,\eta) = (0,0)$, $(0.05, 0)$ and $(0.05,
  0.0721)$, respectively. } 
\label{fig:PolarizationBIC2}
\end{figure}

The BIC S2 is a scalar $E$-polarized propagating BIC. Our numerical
results confirm that it can be extended to the $\delta$-$\eta$ plane
on a curve,  as shown in Fig.~\ref{fig:BIC2}(a). 
The extended BIC
remains scalar (i.e., $\alpha=0$) for all $\delta$. The  wavenumber $\beta$
and frequency $\omega$ of this BIC (for different $\delta$) are shown
in Fig.~\ref{fig:BIC2}(b) as the solid blue and dashed red curves, respectively. 
For $\delta = 0.05$ and $\eta = 0.0721$, we obtain a BIC
with $\beta = 0.2223 (2\pi/L)$
and $\omega = 0.6307 (2\pi c/L)$.
The electric field  magnitude (i.e.  $ |E_x| $) is shown in Fig.~\ref{fig:BIC2}(c).   In
Fig.~\ref{fig:BIC2}(d), we show the $Q$ factor of resonant modes for 
$\eta$  near $0.0721$ and $\beta$ near $0.2223 (2\pi/L)$.
In Fig.~\ref{fig:PolarizationBIC2},
we show polarization ellipses
(upper panels) and polarization directions (lower panel) for resonant
modes near the BIC and the CPSs. 
Keeping $\eta=0$ and increasing $\delta$ to $0.05$, the BIC S2 splits 
into two CPSs with topological charges $+1/2$ as shown in 
Fig.~\ref{fig:PolarizationBIC2}(b). Keeping $\delta=0.05$ and
increasing  $\eta$ to $0.0721$, the two CPSs merge and form a BIC with
$\beta = 0.2223 (2\pi/L)$ and $\omega = 0.6307 (2\pi c/L)$, 
as shown in Fig.~\ref{fig:PolarizationBIC2}(c).

\begin{figure}[htp]
\centering 
\includegraphics[scale=1]{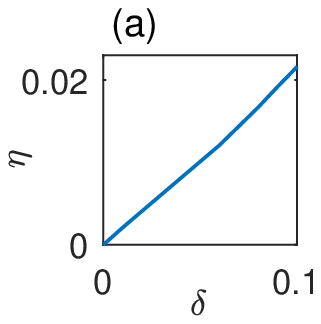}
\includegraphics[scale=1]{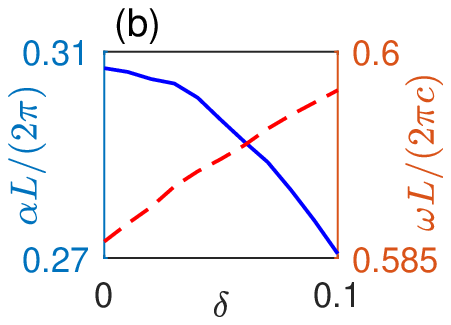}
\includegraphics[scale=1]{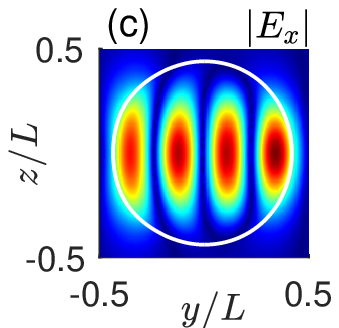}
\includegraphics[scale=1]{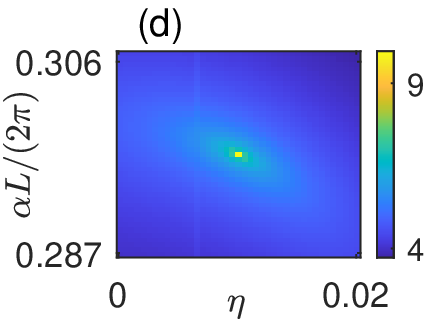}
\caption{BIC near V1 in a perturbed array of cylinders.  (a):
  parameter $\eta$ of the BIC as a function of $\delta$. (b):
  wavenumber $\alpha$ (solid blue curve, left axis)  and frequency
  $\omega$ (dashed red curve, right axis) of the BIC as functions of
  $\delta$. (c): magnitude  of the BIC (i.e. $|{E}_x|$) for $\delta =
  0.05$,  $\eta = 0.0101$, $\alpha=0.2968 (2\pi/L)$, $\beta = 0$, and
  $\omega = 0.5922 (2\pi c/L)$. (d): logarithmic value of the $Q$
  factor (i.e. $\log_{10}Q$) of resonant modes as a function of $\eta$
  and $\alpha$ for $\beta=0$ and $\delta = 0.05$. } 
\label{fig:BIC3}
\end{figure}

\begin{figure}[htp]
\centering 
\includegraphics[scale=0.85]{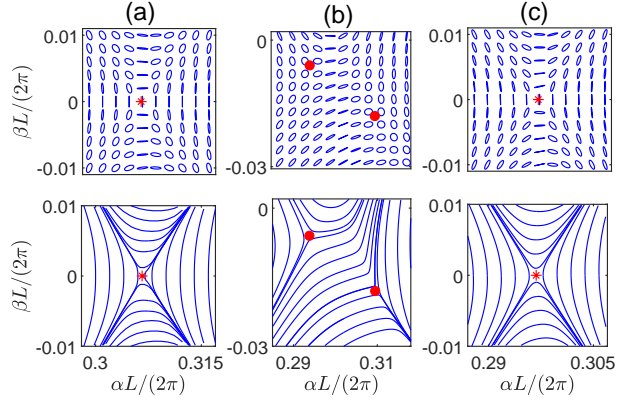}
\caption{Emergence and annihilation of CPSs near BIC V1. (a), (b) and
  (c): polarization ellipses (top panels) and polarization directions
  (lower panels) for  $(\delta,\eta) = (0,0)$, $(0.05, 0)$ and $(0.05,
  0.0101)$, respectively. } 
\label{fig:PolarizationBIC3}
\end{figure}

The BIC V1 is a vectorial BIC propagating along the $x$ axis. Our
numerical results, shown in Fig.~\ref{fig:BIC3}(a),
confirm that this
BIC exists continuously on a curve in the $\delta$-$\eta$ plane. The
wavenumber $\beta$ of this BIC remains at zero for all $\delta$. 
The wavenumber $\alpha$ and frequency $\omega$ are shown 
as functions of $\delta$ in Fig.~\ref{fig:BIC3}(b).
For  $\delta = 0.05$, the BIC is obtained with $\eta = 0.0101$, 
$\alpha=0.2968 (2\pi/L)$, and $\omega = 0.5922 (2\pi c/L)$. The
magnitude of $E_x$ of this BIC is shown in Fig.~\ref{fig:BIC3}(c). 
In Fig.~\ref{fig:BIC3}(d), we show the $Q$ factor of resonant modes
for nearby values of $\eta$ and $\alpha$ (for fixed $\delta=0.05$ and
$\beta=0$).
Similar to the previous examples and as shown in
Fig.~\ref{fig:PolarizationBIC3},
BIC V1 splits to a pair of CPSs
when $\eta$ is fixed at $0$ and $\delta$ is increased to $0.05$, and
the CPSs merge to a BIC when $\delta$ is fixed at $0.05$ and $\eta$ is
increased to to $ 0.0101$.  
%   Note that these two CPSs do not have the property as the CPSs near
%   BIC 1 and BIC 2. Generally, to make the two CPSs merge at one
%   point in the $\beta$-$\gamma$ plane, we have to tune $\delta$ and
%   $\eta$ simultaneously. Our perturbation theory developed in
%   Sec.~\ref{sec:theory} predicates that for $\delta = 0.05$, there
%   exists a $\eta$ such that the perturbed structure has a BIC. That
%   means the two CPSs can merge to generate a BIC by tuning $\eta$
%   only. It is indeed true. Fixing $\delta=0.05$ and tuning $\eta$ 

\begin{figure}[thp]
\centering 
\includegraphics[scale=0.78]{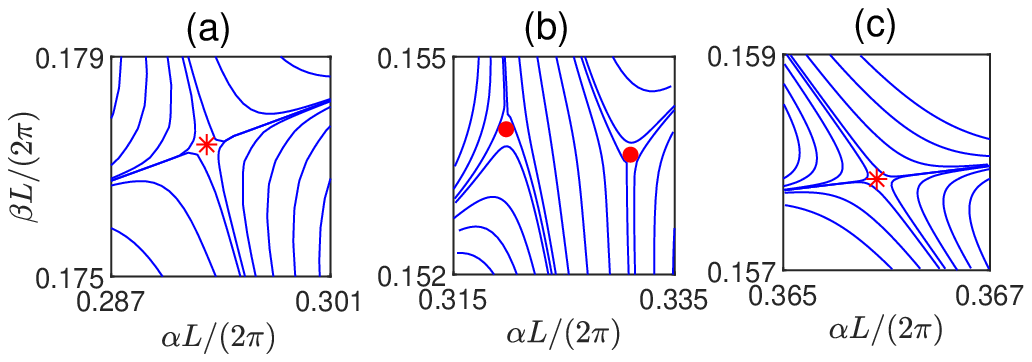}
\includegraphics[scale=1]{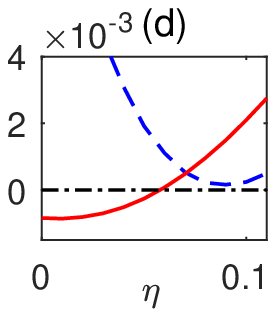}
\includegraphics[scale=1]{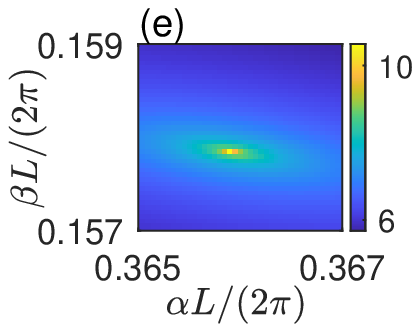}
\includegraphics[scale=1]{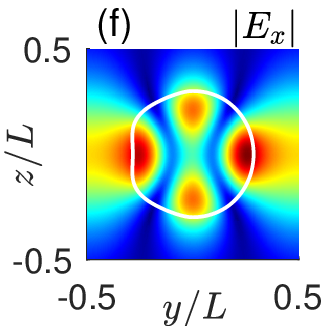}
\caption{Emergence and annihilation of CPSs near BIC V2. (a), (b) and
  (c): polarization directions  for  $(\delta,\eta) = (0,0)$, $(0.05,
  0)$, and $ (0.0418, 0.0556)$, respectively. Red asterisks and dots
  denote BICs and CPSs, respectively. (d): the difference of $\alpha$
  (dashed blue curve) and $\beta$ (solid red curve) of the two CPSs
  as functions of $\eta$ for fixed $\delta=0.05$. The unit of the
  vertical axis is $2\pi/L$. The horizontal dash-dot line denotes
  zero. (e): the logarithmic value of the $Q$ factor (i.e. $
  \log_{10}Q $) of nearby resonant modes as a function of $\alpha$
  and $\beta$ for $(\delta, \eta) = (0.0418, 0.0556)$. There is a BIC
  with $(\alpha, \beta) = (0.3659,0.1578) (2\pi/L)$ and $\omega =
  0.6303 (2\pi c/L)$.  (f): magnitude of the BIC (i.e. $|{E}_x|$).} 
\label{fig:PolarizationBIC4}
\end{figure}

The BIC V2 is a vectorial BIC with both $\alpha_*\ne 0$ and $\beta_*\ne
0$. According to our theory, it can exist continuously on a curve
in a 3D parameter space of $\delta$, $\eta$ and $s$. 
In general, we can only expect the curve to have discrete
intersections with the $\delta$-$\eta$  plane. 
Therefore, the BIC may exist at isolated points in
the plane. To find the BIC with $(\delta, \eta) \ne (0,0)$, we can vary
the parameters and try to force the CPSs to merge. 
In  Fig.~\ref{fig:PolarizationBIC4},
we show the polarization directions of the nearby resonant modes  and
the BIC or CPSs,  for $(\delta,\eta) =(0,0)$, $(0.05, 0)$, and
$(0.0418, 0.0556)$ in panels (a), (b) and (c), respectively. 
For a fixed $\eta
= 0$ and an increasing $\delta$, BIC V2
splits to a pair of CPSs. At $\delta=0.05$ and $\eta=0$, the
wavevectors of the two CPSs are
$(\alpha_1, \beta_1) = (0.3198, 0.1543) (2\pi/L)$ and $(\alpha_2,
\beta_2) = (0.3308, 0.1535) (2\pi/L)$, 
respectively, and are shown in Fig.~\ref{fig:PolarizationBIC4}(b) as
the red dots.
In Fig.~\ref{fig:PolarizationBIC4}(d), we show $\alpha_1-\alpha_2$ and $\beta_1-\beta_2$ 
 of the two CPSs for fixed $\delta = 0.05$ and 
different $\eta$. Since the two curves do not intersect at zero, 
the two CPSs do not merge (become a  
BIC) if $\eta$ is varied and $\delta$ is fixed at $0.05$. 
However, these two CPSs do merge to a BIC if $(\delta, \eta)$ is tuned
to $(0.0418, 0.0556)$. The wavevector of this BIC is $(\alpha, \beta) =
(0.3659, 0.1578) (2\pi/L)$ as shown in  
Fig.~\ref{fig:PolarizationBIC4}(c). Its frequency is
$\omega = 0.6303 (2\pi c/L)$ and its topological charge is $-1$. 
In Fig.~\ref{fig:PolarizationBIC4}(e), we show the $Q$ factor of nearby
resonant modes for $(\delta, \eta) = (0.0418, 0.0556)$. The divergence
of the $Q$ factor  confirms the existence of this BIC. The magnitude
of $E_x$ of this BIC is shown in Fig.~\ref{fig:PolarizationBIC4}(f).

\section{Conclusion}
\label{sec:conclusion}

Dielectric periodic structures sandwiched between two homogeneous
media are a popular platform for studying photonic BICs and their
applications. We analyzed how the BICs in 2D structures with an up-down
mirror symmetry and a single periodic direction depend on structural parameters, 
and showed that a typical vectorial BIC with nonzero
wavenumber components in both invariant and periodic directions
can only be found by tuning two structural parameters, while the other
BICs, including the scalar ones studied in our previous
work~\cite{yuan20_2}, can be found by tuning one structural
parameter. The theory is applicable to generic BICs with a low
frequency such that there is only one radiation channel and for
structures without a reflection symmetry in the periodic
direction. The result is somewhat unexpected, since all these BICs
exhibit the same robustness when the structure and the perturbation
is symmetric in the periodic direction~\cite{yuan17ol,yuan21}.
Our theory is established using an all-order perturbation method and
validated by numerical examples involving periodic arrays of
dielectric cylinders. The numerical studies also demonstrate the 
annihilation and generation of BICs and CPSs as structural parameters
are varied, while the net topological charge remains unchanged. 

Our study suggests that a typical BIC can exist continuously when
a sufficient number of structural parameters are introduced. The
number of parameters depends on the number of opening diffraction 
channels, the retained symmetries, and the nature of the BIC. Although
our theory is developed for 2D structures with a single periodic
direction, it can be extended to 3D structures with two-dimensional
periodicity, and similar results are expected.

\section*{Acknowledgement}
 The authors acknowledge support from the Natural Science Foundation
 of Chongqing, China (Grant No. cstc2019jcyj-msxmX0717),  the program for the Chongqing Statistics Postgraduate Supervisor Team (Grant No. yds183002), and the
 Research Grants Council of Hong Kong Special Administrative Region,
 China (Grant No. CityU 11305518). 

\section*{Appendix}
\renewcommand{\theequation}{A\arabic{equation}}
\setcounter{equation}{0}
%
%The operator $\mathcal{B}$ in Eq.~(\ref{eq:a22Zero}) is defined as
%\begin{equation}
%\mathcal{B} =  i \left[ (\nabla + i \gamma {\bm e}_3) \times {\bm e}_3 \times \cdot + {\bm e}_3 \times (\nabla + i \gamma {\bm e}_3) \times \cdot \right].
%\end{equation}
%Equation (\ref{eq:a22Zero}) implies $\int_{\Omega} \overline{{\bm \Psi}}_2 \cdot \mathcal{B}_2 {\bm \Phi}_* d {\bm r} = 0$ where operator $\mathcal{B}_2$ is defined below.

The governing equations for vector function ${\bm \Phi}({\bm r})$
defined in Eq.~(\ref{eq:Bloch}) are 
\begin{eqnarray}
  \label{eq:MaxEq3}
&& \left( \nabla + i  {\bm \alpha} \right)  \times  \left(
   \nabla + i  {\bm \alpha} \right)  \times {\bm \Phi} - k^2
   \varepsilon({\bm r}) {\bm \Phi} = 0, \\
&& \label{eq:MaxEq4}
  \left( \nabla + i  {\bm \alpha} \right) \cdot
   [ \varepsilon({\bm r}) {\bm \Phi} ] = 0,
  \end{eqnarray}
  where ${\bm \alpha } = (\alpha, \beta, 0)$.
Substituting expansions (\ref{peps3})-(\ref{series5})  into
Eqs.~(\ref{eq:MaxEq3}) and (\ref{eq:MaxEq4}), and comparing the
coefficients of $\delta^j$  for $j \geq 1$, we obtain
Eqs.~(\ref{eq:Phij}) and (\ref{eq:Phij2}) for
${\bm \Phi}_j$, 
%\begin{eqnarray}
%\label{eq:PhiJ} \mathcal{L} {\bm \Phi}_j &=& \beta_j \mathcal{B}_1 {\bm \Phi}_* + \gamma_j \mathcal{B}_2 {\bm \Phi}_*  + 2 k_* k_j \varepsilon_* {\bm \Phi}_*  \nonumber \\
%                                                    & &  +  k_*^2 \eta_j F({\bm r}) {\bm \Phi}_* + k_*^2 s_j P({\bm r}) {\bm \Phi}_* + {C}_j, 
%\end{eqnarray}
%Operators $\mathcal{L}, \mathcal{B}_n$ for $n=1,2$, $\mathcal{D}_m $ for $0 \leq m < j $ and functions $ {\bm C}_j, W_j, V_m$  are defined as
where
\begin{eqnarray*}
\mathcal{L} &=& (\nabla + i  {\bm \alpha}_*) \times (\nabla + i  {\bm \alpha}_*) \times \cdot - k_*^2 \varepsilon({\bm r}), \\
\mathcal{B}_n &=& - i \left[ (\nabla + i {\bm \alpha}_*) \times {\bm
                  e}_{n} \times \cdot + {\bm e}_{n} \times (\nabla + i
                  {\bm \alpha}_*) \times \cdot \right]
\end{eqnarray*}
for $n=1$ and 2, % ${\bm \alpha}_* = (\alpha_*, \beta_*, 0)$,  
${\bm e}_1 $ and ${\bm e}_2$ are the unit vectors in the $x$ and $y$ directions, 
respectively. The other scalar and vector functions in these two
equations are 
\begin{eqnarray*}
  h_1 &=& -i \varepsilon_* {\bm e}_1 \cdot {\bm \Phi}_*, \\
  h_2 &=& -i \varepsilon_* {\bm e}_2 \cdot {\bm \Phi}_*, \\
  h_3 & = & - (\nabla + i {\bm \alpha}_*)\cdot (F {\bm \Phi}_*), \\
  h_4  & =&  - (\nabla + i {\bm \alpha}_*)\cdot (P {\bm \Phi}_*), \\  
g_j & = & -\nabla \cdot (G {\bm \Phi}_{j-1}) - i G {\bm \alpha}_{j-1} \cdot {\bm \Phi}_* \\
&-& i \sum_{m=1}^{j-1} (\varepsilon_* {\bm \alpha}_m + G {\bm \alpha}_{m-1}) \cdot {\bm \Phi}_{j-m} \\
& - & i \sum_{m=1}^{j-1} (\eta_m F + s_m P) {\bm \alpha}_{j-m}\cdot {\bm \Phi}_* \\
&-& i \sum_{m=1}^{j-1} \sum_{n=1}^m (\eta_n F + s_n P) {\bm
    \alpha}_{m-n} \cdot {\bm \Phi}_{j-m}, \\
  {\bm C}_j &=& {W}_j {\bm \Phi}_* + \sum_{m=1}^{j-1} ( {V}_m {\bm \Phi}_{j-m} + \mathcal{D}_m {\bm \Phi}_{j-m}  \\
           & &   + {\bm \alpha}_m \times {\bm \alpha}_{j-m} \times {\bm \Phi}_* ), \\
{V}_m & =& W_m + 2 k_* k_m \varepsilon_* + k_*^2 \eta_m F + k_*^2 s_m P,  
\end{eqnarray*}
\begin{eqnarray*}      
{W}_j &=& \sum_{m=1}^{j-1} \left[ k_m k_{j-m} \varepsilon_* + \sum_{l=0}^{m} k_l k_{m-l} ( \eta_{j-m} F + s_j P ) \right]  \\ 
 & & + \sum_{m=0}^{j-1} k_m k_{j-1-m} G, \\
  \mathcal{D}_m &=& \sum_{l=1}^{m-1} {\bm \alpha}_l \times {\bm \alpha}_{m-l} \times \cdot  \\ 
&-& i [ (\nabla + i {\bm \alpha}_*) \times {\bm \alpha}_m \times \cdot
    + {\bm \alpha}_m \times (\nabla + i {\bm \alpha}_*) \times \cdot
    ], 
\end{eqnarray*}
where ${\bm \alpha}_m = (\alpha_m, \beta_m, 0)$, $k_0 = k_*$,
${\bm \Phi}_0 = {\bm \Phi}_*$, ${\bm \alpha}_0 = {\bm \alpha}_*$, 
and $ \eta_0 = s_0 = 0$. More specifically,  for $j=1$, we have $C_1 = k_*^2 G
{\bm \Phi}_*$. Note that ${\bm \Phi}_*$ satisfies $\mathcal{L} {\bm
  \Phi}_* = 0$. 

The entries $a_{1m}$ for $m=1,2,3,4$ in the first row of coefficient
matrix ${\bm A}$ of linear system (\ref{3by5}) are defined as 
 \begin{eqnarray*}
 a_{11} &=& \int_{\Omega}  \overline{{\bm \Phi}}_* \cdot \mathcal{B}_1
            {\bm \Phi}_* d {\bm r}, \\
   a_{12} &=& \int_{\Omega}  \overline{{\bm \Phi}}_* \cdot \mathcal{B}_2 {\bm \Phi}_* d {\bm r}, \\
 a_{13} &=& 2 k_* \int_{\Omega} \varepsilon_*({\bm r}) \overline{{\bm \Phi}}_* \cdot {\bm \Phi}_* d {\bm r} = 2 k_* L^2,\\
 a_{14} & = & k_*^2 \int_{\Omega} F({\bm r}) \overline{{\bm \Phi}}_*
              \cdot {\bm \Phi}_* d {\bm r},  \\
   a_{15}  &=&  k_*^2 \int_{\Omega} P({\bm r}) \overline{{\bm \Phi}}_* \cdot {\bm \Phi}_* d {\bm r}.
%a_{21} &=& \int_{\Omega}  \overline{{\bm \Psi}}_1 \cdot \mathcal{B}_1 {\bm \Phi}_* d {\bm r}, \\
% a_{22} &=& \int_{\Omega}  \overline{{\bm \Psi}}_1 \cdot \mathcal{B}_2 {\bm \Phi}_* d {\bm r}, \quad a_{24}  =  k_*^2 \int_{\Omega} F({\bm r}) \overline{{\bm \Psi}}_1 \cdot {\bm \Phi}_* d {\bm r}, \\
% a_{31} &=&  \int_{\Omega}  \overline{{\bm \Psi}}_2 \cdot \mathcal{B}_1 {\bm \Phi}_* d {\bm r},  \quad a_{34} = k_*^2 \int_{\Omega}  F({\bm r}) \overline{{\bm \Psi}}_2 \cdot  {\bm \Phi}_* d {\bm r}, \\
%a_{25}  &=&  k_*^2 \int_{\Omega} P({\bm r}) \overline{{\bm \Psi}}_1 \cdot {\bm \Phi}_* d {\bm r}, \quad a_{35} = k_*^2 \int_{\Omega}  P({\bm r}) \overline{{\bm \Psi}}_2 \cdot  {\bm \Phi}_* d {\bm r},
 \end{eqnarray*}
The entries  $a_{2m}$ and $a_{3m}$ for $m=1,2,3,4$ are defined
similarly as $a_{1m}$ with $\overline{{\bm \Phi}}_*$ replaced by
$\overline{{\bm \Psi}}_1$ and $\overline{{\bm \Psi}}_2$, respectively.  
Condition (\ref{eq:EScatt1_Orth}) implies the $(2,3)$ and $(3,3)$
entries of matrix ${\bm A}$ are zero, i.e. $a_{23} = a_{33} = 0$. 
%Condition (\ref{eq:a22Zero}) give rises to $a_{32} = 0$.  
The elements in the right-hand side vector ${\bm b}_j$ of linear system (\ref{3by5}) are
 \begin{eqnarray*}
 b_{1j} &=& -  \int_{\Omega}  \overline{{\bm \Phi}}_* \cdot  {\bm C}_j
            d {\bm r}, \\
   b_{21} &=& -  \int_{\Omega}  \overline{{\bm \Psi}}_1 \cdot   {\bm C}_j d {\bm r}, \\
\label{bj} b_{31} &=& - \int_{\Omega} \overline{{\bm \Psi}}_2 \cdot   {\bm C}_j d {\bm r}.
 \end{eqnarray*}

The linear system (\ref{3by5}) is a necessary and sufficient condition
for Eq.~(\ref{eq:Phij}) to have a solution that decays to zero
exponentially as $|z| \to \infty$.  
If ${\bm \Phi}_j$ decays to zero exponentially as $|z| \to \infty$, we 
take the dot product of Eq.~(\ref{eq:Phij}) with $\overline{{\bm
    \Phi}}_*$, $\overline{{\bm \Psi}}_1$ and $\overline{{\bm \Psi}}_2$,
respectively, integrate the results on domain $\Omega$ as in Appendix
A of Ref.~\cite{yuan21}, and obtain linear system (\ref{3by5}). 
 On the other hand, if system~(\ref{3by5}) has a real solution, the first
 equation in (\ref{3by5}) ensures that  
Eq.~(\ref{eq:Phij}) is solvable, and the second and third equations of
(\ref{3by5}) guarantee that the solution of Eq.~(\ref{eq:Phij}) decays exponentially to zero
as $|z| \to \infty$.

More specifically,  the inhomogeneous equation (\ref{eq:Phij}) has a non-zero solution only if the right hand
side is orthogonal to ${\bm \Phi}_*$, that is, if the first equation 
of linear system (\ref{3by5}) is true. Since there is only one
opening diffraction channel, the asymptotic formulae of diffraction
solutions ${\bm \Psi}_m$ (for $m=1,2$) at infinity are
\[
  {\bm \Psi}_m \sim {\bm f}_m^{\pm} e^{  \pm i \gamma_* z} +   {\bm g}_m^{\mp} e^{ \mp i
  \gamma_* z}, \quad z \to \mp \infty, 
\]
where ${\bm g}_m^{\pm} =
(g_{mx}, g_{my}, \pm g_{mz} ) $ are constant vectors satisfying ${\bm
  g}_m^{\pm}  \cdot   {\bm k}_*^{\pm}  = 0$ for
${\bm k}_*^{\pm} = (\alpha_*, \beta_*, \pm \gamma_*)$
and $\gamma_* = (k_*^2 \varepsilon_0 - \alpha_*^2 - \beta_*^2)^{1/2}$. 
 For each $j \geq 1$, the asymptotic formula of ${\bm \Phi}_j$  at
 infinity is
 \[
   {\bm \Phi}_j \sim {\bm d}_j^{\pm} e^{\pm i \gamma_* z}, \quad z \to \pm
   \infty,
 \]
 where ${\bm d}_j^{\pm} = (d_{jx}, d_{jy}, \pm d_{jz})$ are constant
 vectors satisfying
 ${\bm d}_j^{\pm}  \cdot  {\bm
  k}_*^{\pm}  = 0$. To show that ${\bm \Phi}_j$ decays exponentially, we
only need to show ${\bm d}_j^{+} = {\bm 0}$.  Taking the dot
product of  Eq.~(\ref{eq:Phij})  with $ \overline{{\bm \Psi}}_m$ (for
$m=1,2$), integrating the results on domain $\Omega_h = \left\{ (x,y)
  | |x| < h , |y| < L/2 \right\} $ for $ h > d$, and following the
same procedure in Appendix B of \cite{yuan21}, we have 
 \begin{equation}
  \lim_{h \to \infty} \int_{\Omega_h} \overline{ {\bm \Phi}}_m \cdot \mathcal{L} {\bm \Phi}_j d {\bm r} = -4 i \alpha L \overline{{\bm g}}_m^+ \cdot {\bm d}_j^+.
  \end{equation}
The second and third equations of the linear system (\ref{3by4}) imply
the left hand side in above equation is zero for $m=1,2$. Therefore,
$\overline{{\bm g}}_m^+ \cdot {\bm d}_j^+ = 0$.  Since $\left\{
  {\bm g}_1, {\bm g}_2, {\bm k}_*^+  \right\} $ is a linearly
independent set, we have ${\bm d}_j^+ = {\bm d}_j^- = {\bm 0}$. 

For a vectorial BIC with $\alpha_* \neq 0$ and $\beta_* = 0$, the
electric field ${\bm E}_* = {\bm \Phi}_*$ and the corresponding
diffraction solutions ${\bm E}^{(s)}_1 = {\bm \Psi}_1$ and ${\bm
  E}^{(s)}_2 = {\bm \Psi}_2$ satisfy condition (\ref{eq:RealE}),
i.e. their $x$ components are pure imaginary and their $y$ and
$z$ components are real.  
It is easy to show that $\mathcal{B}_1 {\bm \Phi}_*$ also satisfies 
condition (\ref{eq:RealE}), the $x$ component of $\mathcal{B}_2 {\bm
  \Phi}_*$ is real, and the $y$ and $z$ components  of $\mathcal{B}_2
{\bm \Phi}_*$ are pure imaginary. 
 Therefore, $a_{22} $ and $a_{32}$ are pure imaginary, and  all other
 elements of ${\bm A}$ are real.  For $j=1$, ${\bm C}_1 = k_*^2 G {\bm
   \Phi}_*$ satisfies condition (\ref{eq:RealE}),  $b_{11}$, $b_{12}$
 and  $b_{13}$ are real.
 If $a_{21} a_{34} - a_{24} a_{31} \ne 0$, the complex linear system
(\ref{3by4}) for $j=1$ has a real solution with $ \beta_1=0$ and
$\alpha_1$, $k_1$, $\eta_1$ satisfying Eq.~(\ref{3by3}).  Therefore,
${\bm \Phi}_1$ can be solved from Eq.~(\ref{eq:Phij}) for $j=1$ and it
decays to zero exponentially as $|z| \to +\infty$. In addition,
since the right hand side of Eq.~(\ref{eq:Phij}) for $j=1$ satisfies
Eq.~(\ref{eq:RealE}),
${\bm \Phi}_1$ also can be scaled to satisfies Eq.~(\ref{eq:RealE}). For
any $j \geq 2$, if $\alpha_m$, $k_m$, $\eta_m$ are real, $\beta_m=0$ and
${\bm \Phi}_m$ satisfies  Eq.~(\ref{eq:RealE}) for $ m < j$, then
${\bm C}_j$ satisfies Eq.~(\ref{eq:RealE}) and $b_{1j}$, $b_{2j}$,
$b_{3j}$ are real. The same reasoning can be used to show that linear
system Eq.~(\ref{3by4}) has a real solution with $ \beta_j=0$ and
$\alpha_j$, $k_j$, $\eta_j$ satisfying Eq.~(\ref{3by3}), and ${\bm
  \Phi}_j$ can be solved from  Eq.~(\ref{eq:Phij})  and satisfies
Eq.~(\ref{eq:RealE}).  

A scalar BIC with $\alpha_* = 0$ and $\beta_* \neq 0$,  is either $E$-
or $H$-polarized. Without loss of generality, we assume the BIC ${\bm
  E}_*$ (i.e. ${\bm \Phi}_*$) is $H$-polarized.  
In that case, we let ${\bm f}_1 = (1, 0, 0)$ and ${\bm f}_2 $ be
orthogonal to ${\bm f}_1$ and ${\bm k}_*^+ $, then ${\bm E}_1^{(s)}$
(i.e. ${\bm \Psi}_1$) is $E$-polarized and ${\bm E}_2^{(s)}$
(i.e. ${\bm \Psi}_2$) is $H$-polarized.
It is easy to show that $\mathcal{B}_1 {\bm \Phi}_*$ is $E$-polarized
and $\mathcal{B}_2 {\bm \Phi}_*$ is $H$-polarized.  
Therefore,  $a_{11} = a_{22} =a_{31}= a_{24} = 0$.  For $j=1$,
${\bm C}_1 = k_*^2 G {\bm \Phi}_*$ is also $H$-polarized and $b_{21} = 0$. 
If $a_{32} \neq 0 $ and $\mbox{Im}(a_{34}/a_{32}) \neq 0 $, then the
linear system (\ref{3by4})  for $j=1$ has a real solution with
$\alpha_1 = 0$ and $\beta_1$, $k_1$, $\eta_1$ satisfying Eq.~(\ref{2by3}).   
Thus, ${\bm \Phi}_1$ can be solved from Eq.~(\ref{eq:Phij}) for $j=1$
and ${\bm \Phi}_j \to 0$ exponentially as $|z| \to +\infty$.  
In addition, since the right hand side of Eq.~(\ref{eq:Phij}) for
$j=1$ is $H$-polarized, ${\bm \Phi}_1$ is also $H$-polarized.  
For any $j \geq 2$, if  $\alpha_m=0$, $\beta_m$, $k_m$, $\eta_m$ are
real and ${\bm \Phi}_m$ is $H$-polarized for $ m < j$, then ${\bm
  C}_j$ is $H$-polarized and $ b_{2j}=0$. Therefore, linear system
(\ref{3by4}) has a real solution with $ \alpha_j=0$ and  $\beta_j$, 
$k_j$, $\eta_j$ satisfying Eq.~(\ref{2by3}), and ${\bm \Phi}_j$ can be
solved from  Eq.~(\ref{eq:Phij})  and is $H$-polarized.

\end{document}